\documentclass[preprintnumbers,article,amsmath,amssymb,floatfix,10pt,prd,onecolumn,
superscriptaddress,nofootinbib]{revtex4}
\usepackage[colorlinks=true, pdfstartview=FitV, linkcolor=blue, citecolor=red, urlcolor=magenta]{hyperref}
\usepackage{bbm}
\usepackage{amsfonts}
\usepackage{mathrsfs}
\usepackage{latexsym}
\usepackage{epsfig}
\usepackage{epstopdf}
\usepackage{epstopdf}
\usepackage{graphicx}
\usepackage{amssymb}
\usepackage{amsmath}
\usepackage{dcolumn}
\usepackage{bm}
\usepackage{color}
\usepackage{comment}
\usepackage{xcolor}
\begin{document}
\title{\bf Tunneling analysis of null aether black hole theory in the background of Newman-Janis
algorithm}

\author{Riasat Ali}
\email{riasatyasin@gmail.com}
\affiliation{Department of Mathematics, GC,
University Faisalabad Layyah Campus, Layyah-31200, Pakistan}

\author{Rimsha Babar}
\email{rimsha.babar10@gmail.com}
\affiliation{Division of Science and Technology, University of Education, Township, Lahore-54590, Pakistan}

\author{Muhammad Asgher}
\email{m.asgher145@gmail.com}
\affiliation{Department of Mathematics, The Islamia
University of Bahawalpur, Bahawalpur-63100, Pakistan}

\author{G. Mustafa} \email{gmustafa3828@gmail.com}
\affiliation{Department of Mathematics, Shanghai University,
Shanghai-200444, Shanghai, People's Republic of China}

\begin{abstract}
We present a new asymptotically flat black hole solution in null aether theory (NAT) by applying Newman-Janis process. For this purpose, we study the asymptotically flat NAT black hole solution in Newman-Janis
algorithm and then compute the tunneling radiation for NAT black hole. The Hawking temperature for
NAT black hole depends upon the rotation parameter and charge of the black hole. The Hawking temperature
describes a black hole with extremal event horizon. Furthermore, we analyze the graphical interpretation of
Hawking temperature w.r.t event horizon and check the stability of black hole under the influence of
different parameters associated with black hole temperature.\\\\{\bf keywords:} Null Aether Black Hole Theory; Newman-Janis algorithm; Quantum gravity; Lagrangian field equation; Hamilton-Jacobi phenomenon; Hawking temperature.
\end{abstract}

\maketitle

\date{\today}


\section{Introduction}
According to general quantum theory of gravity Lorentz symmetry won't hold precisely in nature. Lately, this idea has propelled a lot of intentions in Lorentz breaking theories of gravity. Among these frameworks are specific concepts of vector-tensor speculations with favored direction settled at each point of space-time via vector field with fixed-norm. Therefore, vector-tensor gravity speculations are of physical significance nowadays
since they may reveal some aspects of the inside framework of quantum theory of gravity. Einstein-Aether theory  \cite{21} is one of the main theory
in which the Aether field is supposed to be time-like and thus rests the boost
segment of the Lorentz symmetry. The concept of Aether theory has been explored throughout the years from different aspects \cite{22}.
There additionally showed up some related works \cite{23}-\cite{25} which talk about the plausibility of a space-like Aether
field that breaks the rotational invariance.  The dynamics and inner formalism of these theories are yet under consideration, for instance, the stability issue of the aether field has been studied \cite{26}, obviously, to acquire a clear perceptive in this regard, one additionally desires some specific analytic solutions for the genuinely complex equations of motion which these speculations holds.
Null Aether Theory (NAT) is the new vector-tensor speculations of modified theory of gravity \cite{27}. According to NAT, the dynamical vector field acts like the Aether and the BH solution via charge is conceivable \cite{28} as well as the physical properties (ADM mass, thermodynamics, singularity) of the NAT BH have been investigated. Additionally, NAT charge is capable to decrease the thermodynamics of horizon as that of the Reissner Nordstr\"{o}m-AdS BH and generalize the circular orbits of massive as well as massless particles around the BH. Xu and Wang analyzed \cite{28q} the quintessence field  around the Kerr-Newman-AdS BH solution by using the Newman-Janis method and complex calculations.

A typical element of different quantum gravity speculations, like loop quantum gravity, string theory and geometry of non-commutative, is the presence of a minimum observable length \cite{[a],[b]}. The
generalized uncertainty principle (GUP) is a basic approach to understanding this minimal observable length. The generalized commutation relation can be defined as \cite{[c]}
\begin{equation}
[x,p]=i\hbar\left(1+\rho p^2\right),
\end{equation}
here $\rho=\frac{1}{3M^{2}_f}$ represents the correction parameter and $M_f$ is the Plank's mass. Moreover, the generalized uncertainty relation is given as
\begin{equation}
\Delta x\Delta p\geq \frac{\hbar}{2}\left(1+\rho p^2\right),
\end{equation}
where $x$ and $p$ stands for position and momentum operators, respectively.
The GUP effect on the tunneling radiation of the higher dimensional BHs in the context of the boson phenomenon of the spin-1 particle have been studied in \cite{29a,29b}.
The Kerr Newman-NUT-Kiselev solution of BH by applying Newman-Janis approach to the dyonically as well as electrically charged BH encompassed by quintessence has been examined \cite{10}.
The thermodynamical properties of BH (Temperature, heat capacity, angular momentum and entropy) have also been derived. The Hawking temperature phenomenon for the different spin particles has been widely analyzed in literature \cite{11}-\cite{AP}. Moreover, it has been studied that the Hawking temperature for different spin of particles remain preserved.

The aim of our paper is to study the NAT BH solution in the context of Newman-Janis algorithm and to investigate the NAT BH Hawking temperature ($T_{H}$) under the effects of rotation parameter and to
describe a comparison of our new results with previous literature.
Furthermore, to derive the quantum corrected temperature $T'_{H}$ for NAT BH with rotation parameter accompanying GUP effects and to analyze the stable condition of BH in the presence of quantum gravity effects.

This article is formatted in the following manner: Section \textbf{II}, contains a brief introduction about the metric of
asymptotically flat NAT BH. Section \textbf{III} investigate the Hawking temperature of BH under the influence
of Newman-Janis
algorithm. Both Sec. \textbf{IV} and \textbf{VI}, present the graphical analysis of Hawking temperature
w.r.t event horizon.  Section \textbf{V} study the temperature of NAT BH under the influence of quantum
gravity and rotation parameter. Section \textbf{VII}, comprised the summary and discussion of all the results.

\section{Asymptotically flat BH in Null Aether Theory}
The spacetime for asymptotically flat BH in NAT can be defined as \cite{1a}
\begin{equation}
ds^{2}=-E(r)dt^2+\frac{1}{E(r)}dr^2+r^2 d\theta^2+r^2sin^2\theta d\phi^2,\label{1}
\end{equation}
where
\begin{equation}
E(r)=\left\{\begin{array}{ll}
1-\frac{2 \tilde{a}_{1}^{2} \tilde{b}_{1}}{r^{1+\tilde{q}}}-\frac{2 \tilde{a}_{2}^{2} \tilde{b}_{2}}{r^{1-\tilde{q}}}-\frac{2 \bar{m}}{r} & (\text { when } \tilde{q} \neq 0), \\
1-\frac{2 m}{r} & (\text {when } \tilde{q}=0),
\end{array}\right.
\end{equation}
here $E(r)$ shows the metric function, $\tilde{a}_1, \tilde{a}_2, \tilde{b}_1$
and $\tilde{b}_2$ are (constant null vector denoting the Aether field) integration constants treated as free parameters, also
\begin{equation}
\tilde{b}_1=\frac{1}{8}\left[\tilde{c}_3+\tilde{c}_{23}\tilde{q}-3\tilde{c}_2\right],
~~~\tilde{b}_2=\frac{1}{8}\left[\tilde{c}_3-\tilde{c}_{23}\tilde{q}-3\tilde{c}_2\right],~~~\tilde{q} \equiv \sqrt{9+8\frac{\tilde{c}_1}{\tilde{c}_{23}}},
\end{equation}
$\tilde{q}$ gives the charge, $\tilde{m}$ \& $m$ denotes
the mass parameter and $\tilde{c}_1, \tilde{c}_2, \tilde{c}_3$ represents
the dimensionless constant parameters. For the case $\tilde{q}=0$, we can observe that the metric
converts into the usual asymptotically flat Schwarzschild BH. Although, for the case $\tilde{q}\neq0$, we get different
asymptotically flat boundary conditions by considering the following cases independently(by def $\tilde{q}>0$ \cite{1a}):
\begin{equation}
\left.E(r)\right|_{r \rightarrow \infty}=1\left\{\begin{array}{ll}
\text { for } 0<\tilde{q}<1 & \text { (if } \tilde{a}_{1} \neq 0 \text { and }
\left.\tilde{a}_{2} \neq 0\right) \text { or }\left(\text { if } \tilde{a}_{1}=0 \text { or } \tilde{b}_{1}=0\right), \\
\text { for } 0<\tilde{q} & \text { (if } \tilde{a}_{2}=0 \text { or } \left.\tilde{b}_{2}=0\right).
\end{array}\right.
\end{equation}
For $\tilde{q}=0$, one can obtain the ADM mass as
\begin{equation}
\bar{M}_{ADM}=\frac{m}{\bar{G}},
\end{equation}
where we have defined \cite{RR2}
\begin{equation}
\bar{G}=\frac{G}{1-\tilde{c}_{1}\tilde{b}_{1}^{2}}.\label{112}
\end{equation}
The effective value of Newtonian constant $\bar{G}$ associated to the constant $G$ can be
evaluated through experiments within the solar system \cite{RR1}.
Also, for the case $\tilde{q}\neq0$, we get
\begin{equation}
\bar{M}_{ADM}=\frac{1}{\bar{G}}\left[\tilde{m}+
\frac{\tilde{a}^2_{1}\tilde{b}_1}{r^{\tilde{q}}}(1+\tilde{q})+\frac{\tilde{a}^2_{2}
\tilde{b}_2}{r^{\tilde{-q}}}(1-\tilde{q})\right]\Big|_{r \rightarrow \infty}.
\end{equation}
From the above equations, we can get the ADM mass for the NAT BH in the following form
\begin{equation}
\bar{M}_{ADM}=\frac{\tilde{m}}{\bar{G}}\left\{\begin{array}{ll}
\text { for } 0<\tilde{q}<1 & \text { (if } \tilde{a}_{1} = 0 \text { or }
\left.\tilde{b}_{1} = 0\right), \\
\text { for } 0<\tilde{q} & \text { (if } \tilde{a}_{2}=0 \text { or } \left.b_{2}=0\right).
\end{array}\right.
\end{equation}
If we consider the condition $\tilde{a}_2=0$, then, the the Aether field $\phi(r)$ and $E(r)$ becomes
\begin{eqnarray}
E(r)&=&1-\frac{2\tilde{a}^2_{1}\tilde{b}_1}{r^{(1+\tilde{q})}}-\frac{2\tilde{m}}{r},\label{b1}\\
\phi(r)&=&\frac{\tilde{a}_1}{r^{(1+\tilde{q})/2}}.\label{c1}
\end{eqnarray}
We can get the event horizon $r_{0}$ by considering $E\left(r_{0}\right)=0$ and
the horizon area is $A=4 \pi r_{0}^{2}$ . By taking $\tilde{a}_{1}=\tilde{G}
\tilde{Q} r_{0}^{(\tilde{q}-1) / 2}$ the Eqs. (\ref{b1}) and (\ref{c1}) become
\begin{eqnarray}
E(r)&=&1-\frac{2 \tilde{G}^{2} \tilde{Q}^{2} \tilde{b}_{1}}{r^{2}}
\left(\frac{r_{0}}{r}\right)^{\tilde{q}-1}-\frac{2 \tilde{m}}{r}, \\
\phi(r)&=&\frac{\tilde{G} \tilde{Q}}{r}\left(\frac{r_{0}}{r}\right)^{(\tilde{q}-1) / 2},
\end{eqnarray}
here $\tilde{Q}$ depicts the charge of NAT BH.
After putting $r=r_{0}$ in  the above equations, we get
\begin{eqnarray}
E\left(r_{0}\right)&=&1-\frac{2 \bar{G}^{2} \tilde{Q}^{2}
\tilde{b}_{1}}{r_{0}^{2}}-\frac{2 \tilde{m}}{r_{0}}=0,\label{d1}\\
\phi\left(r_{0}\right)&=&\frac{\bar{G} \tilde{Q}}{r_{0}}.\label{e1}
\end{eqnarray}
It is note worthy to mention here that the horizon condition in Eq. (\ref{d1})
is free from the parameter $\tilde{q}$. Moreover, $\phi(r)$ looks like the electric potential at $r=r_{0}$.

After substituting $\tilde{q}=1$ in the metric (\ref{1}), the $E(r)$ and $\phi(r)$ get the form
\begin{equation}
\begin{array}{l}
E(r)=1-\frac{2 \tilde{a}_{1}^{2} \tilde{b}_{1}}{r^{2}}-\frac{2 \tilde{m}}{r}, \\
\phi(r)=\frac{\tilde{a}_{1}}{r^{1 / 2}}.
\end{array}
\end{equation}
\section{Asymptotically flat NAT BH in Newman-Janis
algorithm}
By applying the Newman-Janis algorithm \cite{30,31, 31a}, we generalize the asymptotically flat NAT BH
solution.
Now we introduce a coordinate transformation from Boyer Lindquist (BL) coordinates
$(t, r, \theta, \phi)$ to Eddington Finkelstein (EF) coordinates $(u, r, \theta, \phi)$
\begin{equation}
du=dt-\frac{dr}{E(r)},\label{A}
\end{equation}
where $u$ represents the null coordinate.
According to new coordinates the Eq. (\ref{1}) can be rewritten as
\begin{equation}
ds^{2}=-E(r)du^2+r^2 d\theta^2-2dudr+r^2 \sin^2\theta d\phi^2.\label{f1}
\end{equation}
The non-zero components for the inverse metric (\ref{f1}) are defined as
\begin{equation}
g^{ur}=-1,~~g^{rr}=E(r),~~g^{\theta\theta}=\frac{1}{r^2},~~g^{\phi\phi}
=\frac{1}{r^2\sin^2\theta}.\nonumber
\end{equation}
Moreover, the inverse metric with complex null tetrad $Z^x=(l^{x}, n^{x}, m^{x}, \bar{m}^{x})$ can be written as
\begin{eqnarray}
g^{xy}=-l^x n^y-l^y n^x+m^x \bar{m}^{y}+m^y \bar{m}^{x}.\label{h1}
\end{eqnarray}
The corresponding components can be defined as
\begin{eqnarray}
l^{x}&=&\delta_{r}^{x},~~~n^{x}=\delta_{u}^{x}-\frac{1}{2} E(r) \delta_{r}^{x},\nonumber\\
m^{x}&=&\frac{1}{\sqrt{2}r} \delta_{\theta}^{x}+\frac{i}{\sqrt{2}r \sin\theta}\delta_{\phi}^{x},\nonumber\\
\bar{m}^{x}&=&\frac{1}{\sqrt{2}r} \delta_{\theta}^{x}-\frac{i}{\sqrt{2}r \sin^2\theta}\delta_{\phi}^{x}.\nonumber
\end{eqnarray}
These null tetrad have orthonormal relation and comply with the accompanying characterizing
conditions, specifically all the vectors satisfy the given relations
\begin{eqnarray}
l_{x} l^{x} &=&n_{x}n^{x}~~=m_{x}m^{x}=\bar{m}_{x}\bar{m}^{x}=0,\nonumber \\
l_{x}m^{x} &=&l_{x}\bar{m}^{x}~~=n_{x}m^{x}~=n_{x}\bar{m}^{x}=0, \nonumber\\
l_{x}n^{x}&=&m_{x}\bar{m}^{x}=1,\nonumber
\end{eqnarray}
By considering the Newman-Janis method, we enable the coordinates to get complex values,
while for real $l^x$ and $n^x$ we are able to consider the given transformation \cite{6},
\begin{eqnarray}
u^{\prime}&=&u-i a \cos \theta,\nonumber \\
r^{\prime}&=&r+i a \cos \theta ,
\end{eqnarray}
here $a$ represents the spin parameter (due to Newman-Janis
algorithm). Furthermore, we consider the transformations from
$E(r)\rightarrow \tilde{E}(r, a, \theta)$ and $\sigma^2=r^2+a^2 \cos^2\theta$, whereas
the null tetrad transforms as vectors in the form
\begin{eqnarray}
l^{x}&=&\delta_{r}^{x},~~~n^{y}=\delta_{u}^{x}-\frac{1}{2}
\tilde{E}(r) \delta_{r}^{x},\nonumber\\
m^{x}&=&\frac{1}{\sqrt{2}r}\left(\delta_{\theta}^{x}+\frac{i}{\sin\theta}\delta_{\phi}^{x}+ia \sin\theta(\delta_{u}^{x}
-\delta_{r}^{x})\right),\nonumber\\
\bar{m}^{x}&=&\frac{1}{\sqrt{2}r}\left(\delta_{\theta}^{x}-\frac{i}{\sin\theta}\delta_{\phi}^{x}-ia \sin\theta(\delta_{u}^{x}
-\delta_{r}^{x})\right).\label{g1}
\end{eqnarray}
By using the Eq. (\ref{h1}) and (\ref{g1}), the
$g^{xy}$ components of non-zero in the EF coordinate can be defined as
\begin{eqnarray}
g^{uu}&=&\frac{a^2\sin^2\theta}{\sigma^2},~~~g^{ur}=g^{ru}=-1-\frac{a^2\sin^2\theta}{\sigma^2},
~~~g^{rr}=\tilde{E}(r, \theta)+\frac{a^2\sin^2\theta}{\sigma^2},~~~
g^{\theta\theta}=\frac{1}{\sigma^2},\nonumber\\
g^{\phi\phi}&=&\frac{1}{\sigma^2\sin^2\theta},~~~g^{u\phi}=g^{\phi u}=\frac{a}{\sigma^2},~~~
g^{r\phi}=g^{\phi r}=-\frac{a}{\sigma^2}.\nonumber
\end{eqnarray}
Furthermore, the lower indices components of matrix in the EF coordinates can be given as
\begin{eqnarray}
g_{uu}&=&-\tilde{E}(r, \theta),~~~g_{ur}=g_{ru}=-1,~~~g_{rr}=0,~~~
g_{\theta\theta}=\sigma^2,~~~g_{u\phi}=g_{\phi u}=a \sin^2\theta,\nonumber\\
g_{\phi\phi}&=&\sin^2\theta\left(\sigma^2+a^2(\tilde{E}(r, \theta)-2)\sin^2\theta\right),~~~
g_{r\phi}=g_{\phi r}=-\frac{a}{\sigma^2},
\end{eqnarray}
where
\begin{equation}
\tilde{E}(r, \theta)=\frac{r^2E+a^2cos^2\theta}{\sigma^2}.
\end{equation}
According to transformed tetrad the new line element can be written as
\begin{eqnarray}
ds^2&=&-\tilde{E}(r,\theta)du^2+\sigma^2d\theta^2+2a\sin^2\theta drd\phi-2a\left(1-\tilde{E}(r,\theta)\right)\sin^2\theta
dud\phi-2dudr\nonumber\\
&+&\sin^2\theta\left(\sigma^2+a^2\left(2-\tilde{E}(r, \theta)\right)\sin^2\theta\right)d\phi^2.
\end{eqnarray}
Now we introduce the transformation of EF coordinates to BL coordinates as \cite{31r}
\begin{equation}
du=dt+Y(r)dr,~~~d\phi=d\phi+\chi(r)dr, \label{j1}
\end{equation}
where the function of $Y(r )$ and $\chi(r )$ is to ignore the $g_{r\phi}$
and $g_{tr}$ components. However, $Y(r )$ and $\chi(r )$ appears as function of $r$ and
$\theta$ which can be defined as

\begin{equation}
Y(r)=-\frac{r^2+a^2}{\left(r^2{E}+a^2\right)},~~~~~~ \chi(r)=-\frac{a}{\left(r^2{E}+a^2\right)}.
\end{equation}
The dependence of $\theta$ from EF to BL coordinates transformation reveals the fact that,
we are dealing with modified theory of gravity and non-vacuum surrounding \cite{31}.
Furthermore, we will exclude the dependency on $r$ and $\theta$ in the functions $\sigma^2$ and $\Delta_r$.
The asymptotically flat NAT BH with BL coordinates in the context of Newman-Janis
algorithm can be obtained as
\begin{eqnarray}
ds^2&=&-\left(\frac{\Delta_r-a^2\sin^2\theta}{\sigma^2}\right)dt^2+\frac{\sigma^2}{\Delta_r} dr^2-2a\left(1+\frac{a^2\sin^2\theta-\Delta_r}{\sigma^2}\right)\sin^2\theta dt d\phi+ \sigma^2 d\theta^2\nonumber
\\&+&\sin^2\theta\left[\sigma^2+\sin^2\theta\left(2-a^2\frac{\Delta_r-a^2\sin^2\theta}{\sigma^2}\right)\right]d\phi^2,\label{k1}
\end{eqnarray}
here
\begin{equation}
\Delta_r=r^2-2mr+a^2-\frac{2 \tilde{a}_{1}^{2} \tilde{b}_{1}}
{\sigma^{2(\tilde{q}-1)/2}}-\frac{2 \tilde{a}_{2}^{2}
\tilde{b}_{2}}{\sigma^{2(-1-\tilde{q})/2}}.
\end{equation}
Since, BH acts like thermodynamical substance and whose temperature $T_{H}$ can be determined
by considering the surface gravity $\kappa$. So, we can compute the Hawking temperature of the
metric (\ref{k1}) by using the following formula \cite{31r}
\begin{eqnarray}
T_{H}=\frac{k}{2\pi},~~~~~k=\frac{\Delta'_r}{2(r^2_+ + a^2)},
\end{eqnarray}
where $\Delta'_r=\frac{d}{dr}(\Delta_r)$. The corresponding Hawking temperature for
NAT BH with Newman-Janis
algorithm can be derived as
\begin{eqnarray}
T_{H}=\left[\frac{r_{+}-m-\tilde{a}^2_1 \tilde{b}_1 r(1-\tilde{q})(r^2_+ + a^2)^{
(-1-\tilde{q})/2}-\tilde{a}^2_2\tilde{b}_2r(1+\tilde{q})(r^2_+ + a^2)^{(\tilde{q}-1)/2}}{2\pi(r^2_+ + a^2)}\right].\label{11}
\end{eqnarray}
The $T_{H}$ for BH depends upon the BH mass $m$, charge $\tilde{q}$, rotation parameter
$a$ and free parameters $\tilde{a}_1, \tilde{a}_2, \tilde{b}_1, \tilde{b}_2$. The above
temperature reduces into temperature of Schwarzschild BH for $a=0,~\tilde{q}=0$ which implies as \cite{32}
\begin{equation}
T_{SBH}=\frac{(r_{+}-m)}{2\pi r^2_+},~~~\text{where}~~~r_{+}=2m.\label{m1}
\end{equation}

\section{Stability Analysis of NAT BH}
This section is comprised to investigate the graphical interpretation of
temperature $T_H$ w.r.t event horizon ($r_{+}$). We evaluate the physical
significance of the plots to analyze the effects of charge $\tilde{q}$, mass $m$ and rotation parameter $a$ of BH on  temperature to study the BH stability.

\textbf{Figure 1}: depicts the presentation of $T_H$ via $r_{+}$
for the fixed values of mass $m=1$, rotation parameter $a=1$, free parameters
$\tilde{a}_1=0.1=\tilde{b}_1, \tilde{a}_2=50, \tilde{b}_2=-10$ in the range of charge $0.1\leq\tilde{q}<0.3$.
At first, the $T_H$ increases and attains a maximum height
and then it drops down gradually from a height and gets an asymptotically flat sate by indicating the stability of BH as $r_{+}\rightarrow\infty$.
It can be observe that the temperature of BH increases
with the decreasing values of horizon. This physical behavior satisfies the Hawking's phenomenon and guarantee the stability of BH.
For $0.1\leq\tilde{q}\leq0.3$, we observe an asymptotically flat behavior in temperature that exhibits the stable state of BH.
\begin{figure}[!tbp]
\centering
\begin{minipage}[b]{0.44\textwidth}
\includegraphics[width=\textwidth]{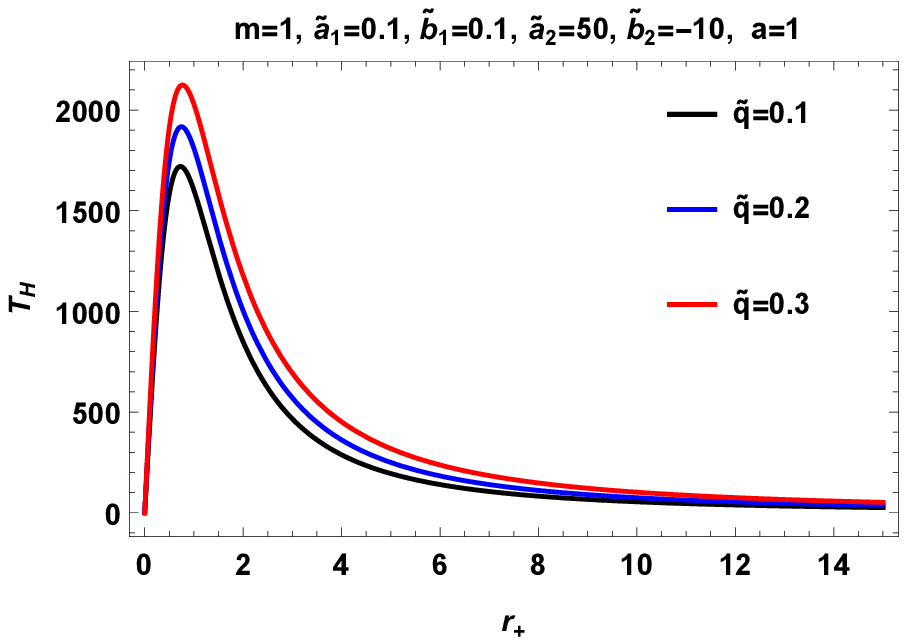}
\caption{$T_{H}$ versus $r_{+}$.}
\end{minipage}
\begin{minipage}[b]{0.44\textwidth}
\includegraphics[width=\textwidth]{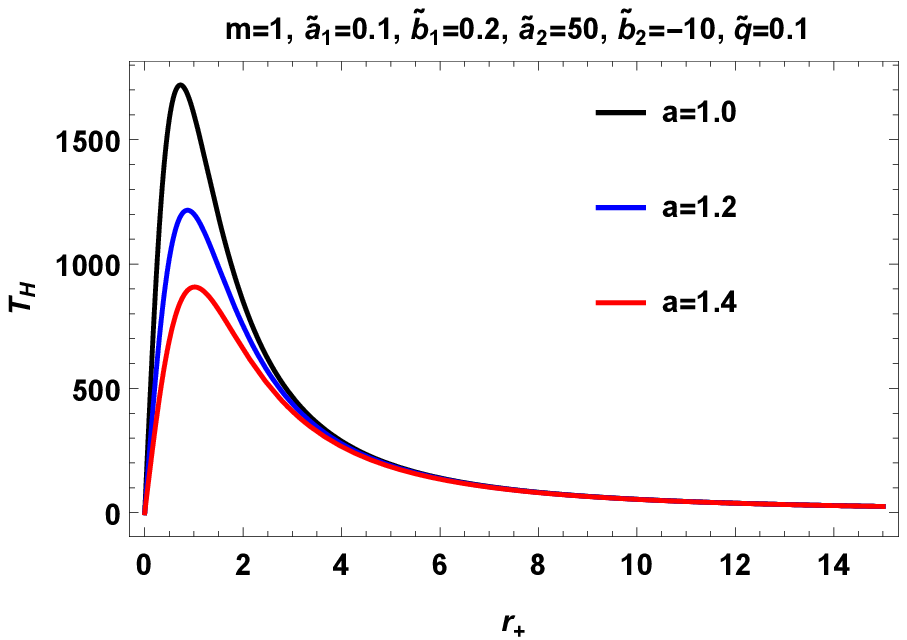}
\caption{$T_{H}$ versus $r_{+}$.}
\end{minipage}
\end{figure}
\textbf{Figure 2}: depicts the behavior of $T_{H}$ via $r_{+}$ with fixed values of mass $m=1$,
charge $\tilde{q}=0.1$, free parameters $\tilde{a}_1=0.1, \tilde{b}_1=0.2, \tilde{a}_2=50,
\tilde{b}_2=-10$ and for varying values of rotation parameter $a$ in the range $0\leq r_+\leq15$.
There can be seen that an asymptotically flat behavior of temperature appears after attaining a
maximum height for different values of $a$ . It can be seen that as we rises the value of $a$ the temperature goes on decreasing as well as for the increasing horizon the
temperature decreases. This Hawking's phenomenon depicts the BH stability in the domain $0\leq r_{+}\leq15$.

It has worth to mention here that for $T_{H}\geq0$, the BH shows the physical behavior and it is in complete stable form.

\section{Temperature of NAT BH Under the Influence of Quantum Gravity}
In this chapter, we analyze the $T_{H}$ under the act upon of quantum gravity for boson spin-$1$ particles.
We rewrite the Eq. (\ref{k1}) in the adopting form
\begin{eqnarray}
ds^{2}&=&-Fdt^{2}+Gdr^{2}+Hd\theta^{2}
+K d\phi^2+2Ldt d\phi,\label{aa}
\end{eqnarray}
where
\begin{eqnarray}
F&=&\frac{\Delta_r-a^2\sin^2\theta}{\sigma^2},
~~~~~~G=\frac{\sigma^2}{\Delta_r},
~~~~~~H=\sigma^2,\nonumber\\
K&=&\sin^2\theta\left[\sigma^2+\left(2+\frac{a^2\sin^2\theta+\Delta_r}{\sigma^2}\right)a^2\sin^2\theta\right]
\nonumber\\L&=&-2a\left(1+\frac{a^2\sin^2\theta+\Delta_r}{\sigma^2}\right)\sin^2\theta.\nonumber
\end{eqnarray}
In order to evaluate the corrected $T_{H}$ of vector particles from the BHs. The vector particles such as $Z$ and $W$ are well-known and act as very significance role in Standard Model \cite{12}. We motion the charges bosonic tunneling in the NAT BH should be more complicated than the Lagrangian field equation as the nontrivial solution interaction during the charged bosonic field, the electromagnetic field and the Aether field. Firstly, we take the field equation of charged particles from the Lagrangian field equation given by the GUP and also we use the Hamilton–Jacobi ansatz phenomenon and WKB approximation to calculate the set of field equation in NAT space-time. By considering the coefficient matrix determinant equal to zero and the linear equations can be derived for the radial function. Accordingly, we compute the tunneling probability of the vector particles from the NAT BH and discuss the corresponding temperature. Therefore, we utilize the generalized Lagrangian equation incorporating the GUP influenced by quantum gravity.
The Lagrangian field equation is given \cite{19} by
\begin{eqnarray}
&&\partial_{\mu}(\sqrt{-g}\chi^{\nu\mu})+\sqrt{-g}\frac{m^2}{\hbar^2}
\varphi^{\nu}+\sqrt{-g}\frac{i}{\hbar}A_{\mu}\varphi^{\nu\mu}+\sqrt{-g}\frac{i}
{\hbar}eF^{\nu\mu}\varphi_{\mu}+\varrho\hbar^{2}\partial_{0}\partial_{0}
\partial_{0}(\sqrt{-g}g^{00}\varphi^{0\nu})\nonumber\\
&&-\varrho \hbar^{2}\partial_{i}\partial_{i}\partial_{i}(\sqrt{-g}g^{ii}\varphi^{i\nu})=0,\label{xx}
\end{eqnarray}
here determinant of $g$, $\varphi^{\nu\mu}$ and $m$ represent coefficient matrix, anti-symmetric
of tensor and particle of mass, since
\begin{eqnarray}
\varphi_{\nu\mu}&=&(1-\varrho{\hbar^2\partial_{\nu}^2})\partial_{\nu}\varphi_{\mu}-
(1-\varrho{\hbar^2\partial_{\mu}^2})\partial_{\mu}\varphi_{\nu}+
(1-\varrho{\hbar^2\partial_{\nu}^2})\frac{i}{\hbar}eA_{\nu}\varphi_{\mu}
-(1-\varrho{\hbar^2}\partial_{\nu}^2)\frac{i}{\hbar}eA_{\mu}\varphi_{\nu},\nonumber\\
F_{\nu\mu}&=&\nabla_{\nu} A_{\mu}-\nabla_{\mu} A_{\nu},\nonumber
\end{eqnarray}
where $\varrho,~A_{\mu}$, $\nabla_{\mu}$ and $~e~$ represent the GUP(quantum gravity) parameter,
vector potential, covariant derivatives and the charge of particle, respectively. The elements of non-zero
for anti-symmetric tensor can be calculated as
\begin{eqnarray}
&&\varphi^{0}=\frac{-K\varphi_{0}+L\chi_{3}}{FK+L^2},~~~\varphi^{1}=\frac{1}{G}\varphi_{1},
~~~\varphi^{2}=\frac{1}{H}\varphi_{2},~~~
\varphi^{3}=\frac{L\varphi_{0}+F\chi_{3}}{FK+L^2},~~\varphi^{12}=\frac{1}{GH}\varphi_{12},
~\varphi^{13}=\frac{1}G{FK+L^2}\varphi_{13},\nonumber\\
&&\varphi^{01}=\frac{-K\varphi_{01}+L\varphi_{13}}{G(FK+L^2)},~~~
\varphi^{02}=\frac{-K\varphi_{02}}{H(FK+L^2)},
~~~\varphi^{03}=\frac{(-FK+F^2)\varphi_{03}}{(FK+L^2)^2},
~~\varphi^{23}=\frac{L\varphi_{02}+F\varphi_{23}}{H(FK+L^2)},\nonumber
\end{eqnarray}
The WKB approximation can be expressed as
\begin{equation}
\varphi_{\nu}=c_{\nu}\exp[\frac{i}{\hbar}Q_{0}(t,r,\phi,\theta)+
\Sigma \hbar^{n}Q_{n}(t,r, \phi,\theta)].
\end{equation}
Using variables technique of separation, we can choose
\begin{equation}
Q_{0}=-\tilde{E}t+W(r)+\nu(\phi)+J\theta,
\end{equation}
where $\tilde{E}=E-J\omega$ and $E$, $J$ denote the particle energy and the angular
particle momentum corresponding to $\theta$ angle. After substituting Eq. (\ref{xx}) into
set of the field equations, we get a matrix of order $4\times 4$
\begin{equation}
Y(c_{0},c_{1},c_{2},c_{3})^{T}=0,
\end{equation}
whose elements are given as follows:
\begin{eqnarray}
Y_{00}&=&\frac{-K}{G(FK+L^2)}\Big[W_{1}^2+\varrho W_{1}^4\Big]
-\frac{K}{H(FK+L^2)}\Big[\nu_{1}^2+\varrho \nu_{1}^4\Big],
-\frac{FK}{(FK+L^2)^2}\Big[J^2+\varrho J^4\Big]-\frac{m^2 K}{(FK+L^2)},\nonumber\\
Y_{01}&=&\frac{\tilde{-K}}{G(FK+L^2)}\Big[L+\varrho
\tilde{E}^3+eA_{0}+\varrho eA_{0}\tilde{E}^2\Big]W_{1}
+\frac{E}{G(FK+L^2)}+\Big[\nu_{1}+\varrho \nu_{1}^3\Big],\nonumber\\
Y_{02}&=&\frac{-K}{H(FK+L^2)}\Big[\tilde{E}+\varrho
\tilde{E}^3-eA_{0}-\varrho eA_{0}\tilde{E}^2\Big]J,\nonumber\\
Y_{03}&=&\frac{-\tilde{E}}{B(FK+L^2)}\Big[W_{1}^2+\varrho W_{1}^4\Big]
-\frac{FK}{H(FK+L^2)^2}\Big[\tilde{E}+\varrho \tilde{E}^3
-eA_{0}-\varrho eA_{0}\tilde{E}^2\Big]J+\frac{m^2L}{(FK+L^2)^2},\nonumber\\
Y_{11}&=&\frac{\tilde{-K}}{G(FK+L^2)}\Big[\tilde{E}^2
+\varrho\tilde{E}^4-eA_{0}\tilde{E}-\varrho eA_{0}\tilde{E}W_{1}^2\Big]
+\frac{L}{G(FK+L^2)}-\frac{m^2}{G}\nonumber\\
&+&\Big[J+\varrho J^3\Big]\tilde{E}-\frac{1}{GH}\Big[\nu_{1}^2+\varrho \nu_{1}^4\Big]-
\frac{1}{G(FK+L^2)}\Big[J+\varrho J^3\Big]+\frac{eA_{0}L}{G(FK+L^2)}\Big[J+
\varrho J^3\Big]\nonumber\\
&-&\frac{eA_{0}K}{G(FK+L^2)}\Big[\tilde{E}+\varrho \tilde{E}^3-eA_{0}
-\varrho eA_{0}\tilde{E}^2\Big],~~~~~~Y_{12}=\frac{1}{GH}[W_{1}+\varrho W_{1}^3]\nu_{1},\nonumber\\
Y_{13}&=&\frac{\tilde{-E}}{G(FK+L^2)}\Big[W_{1}+\varrho W_{1}^3\Big]\tilde{E}
+\frac{1}{G(FK+L^2)^2}\Big[W_{1}+\varrho W_{1}^3\Big]J
+\frac{LeA_{0}}{G(FK+L^2)}\Big[W_{1}+\varrho W_{1}^3\Big],\nonumber\\
Y_{22}&=&\frac{K}{H(FK+L^2)}\Big[\tilde{E}^2
+\varrho \tilde{E}^4-eA_{0}\tilde{E}-\varrho eA_{0}\tilde{E}\Big]
-\frac{1}{GH}-\frac{m^2}{H}\nonumber\\
&-&\frac{F}{H(FK+L^2)}\Big[\nu_{1}^2+\varrho \nu_{1}^4\Big]-\frac{eA_{0}K}{H(FK+L^2)}
\Big[\tilde{E}+\varrho \tilde{E}^3-eA_{0}-\varrho eA_{0}\tilde{E}^2\Big]\nonumber\\
&+&\frac{L}{H(FK+L^2)}\Big[\tilde{E}
+\varrho \tilde{E}^3-eA_{0}-\varrho eA_{0}\tilde{E}^2\Big]J,\nonumber\\
Y_{23}&=&\frac{F}{G(FK+L^2)}\Big[\nu_{1}+\varrho \nu_{1}^3\Big]J
\end{eqnarray}
\begin{eqnarray}
Y_{33}&=&\frac{(FK-\tilde{F^2})}{(FK+L^2)}\Big[\tilde{E}^2
+\varrho \tilde{E}^4-eA_{0}\tilde{E}-\varrho eA_{0}\tilde{E}^3\Big]
-\frac{1}{G(FK+L^2)}\Big[W_{1}^2+\varrho W_{1}^4\Big]\nonumber\\
&-&\frac{F}{H(FK+L^2)}\Big[\nu_{1}^2+\varrho \nu_{1}^4\Big]
-\frac{m^2 F}{(FK+L^2)}-\frac{eA_{0}(FK-\tilde{F^2})}{(FK+L^2)}\Big[\tilde{E}
+\varrho \tilde{E}^3-eA_{0}\tilde{E}^2\Big],\nonumber
\end{eqnarray}
where $\nu_{1}=\partial_{\phi}Q_{0}$, $W_{1}=\partial_{r}Q_{0}$ and $J=\partial_{\theta}{Q_{0}}$.
The determinant of $Y$ is equal to
zero for the non-trivial solution and get
\begin{eqnarray}\label{a1}
ImW^{\pm}&=&\pm \int\sqrt{\frac{(E-J\omega-A_{0}e)^{2}
+X_{1}\Big[1+\varrho\frac{X_{2}}{X_{1}}\Big]}{(FK+L^2)/GK}}dr,\nonumber\\
&=&\pm \pi\frac{(\tilde{E}-A_{0}e)}{2k(r_{+})}\Big[1+\varrho\Xi\Big],
\end{eqnarray}
where
\begin{eqnarray}
X_{1}&=&\frac{GL}{(FK+L^2)}\Big[\tilde{E}
-eA_{0}\Big]\nu_{1}+\frac{FG}{(FK+L^2)}J^2-Gm^2,\nonumber\\
X_{2}&=&\frac{GK}{(FK+L^2)}\Big[\tilde{E}^4-2eA_{0}\tilde{E}^3+(eA_{0})^2
\tilde{E}^2\Big]-\frac{FG}{(FK+L^2)}J^4-W_{1}^4\nonumber\\
&+&\frac{GL}{H(FK+L^2)}\Big[\tilde{E}^3-eA_{0}\tilde{E}^2\Big]J.\nonumber
\end{eqnarray}
The bosonic particle tunneling can be expressed as
\begin{equation}
\Gamma=\frac{\Gamma_{\textmd{emission}}}{\Gamma_{\textmd{absorption}}}=
\exp\left[{-2\pi}\frac{(E-J\omega-A_{0}e)}
{k(r_{+})}\right]\Big[1+\varrho\Xi\Big].
\end{equation}
where
\begin{equation}
k=\frac{\Delta'_r}{2(r^2_+ + a^2)}.
\end{equation}
The modified temperature can be calculated by applying the Boltzmann factor
$\Gamma_{B}=\exp\left[(E-J\omega-A_{0}e)/T'_{H}\right]$ as
\begin{equation}
T'_{H}=\left[\frac{r_{+}-m-\tilde{a}^2_1 \tilde{b}_1 r(1-\tilde{q})(r^2_+ + a^2)^{
(-1-\tilde{q})/2}-\tilde{a}^2_2\tilde{b}_2r(1+\tilde{q})(r^2_+ + a^2)^{(\tilde{q}-1)/2}}{2\pi(r^2_+ + a^2)}\right]\Big[1-\varrho\Xi\Big]\label{thh}.
\end{equation}
The Hawking temperature for BH depends upon the mass $m$, charge $\tilde{q}$, quantum gravity parameter $\varrho$, spin parameter
$a$, arbitrary parameter $\Xi$ and free parameters $\tilde{a}_1$, $\tilde{a}_2$, $\tilde{b}_1$, $\tilde{b}_2$. The
expression (\ref{thh}) reduces into BH temperature for $\varrho=0$, which leads a temperature in Eq. (\ref{11}).
It has note worthy that the quantum corrections cause a deceleration in the increment of temperature.

\section{Stability Analysis of NAT BH with quantum corrections}
This section depicts the graphical presentation of $T'_H$ w.r.t event horizon ($r_{+}$) with fixed value of arbitrary parameter $\Xi=1$. We study the physical
existence of the plots and observe the effects of correction parameter $\varrho$ and spin
parameter $a$ of BH on corrected Hawking temperature to study the stable BH condition under the influence of quantum effects.

\textbf{Figure 3(i)}: describes the behavior of $T'_H$ via event horizon
for the fixed values of mass $m=1$, spin parameter $a=1$, free parameters
$\Xi=1$, $\tilde{a}_1=0.1=\tilde{b}_1$, $\tilde{a}_2=50$, $\tilde{b}_2=-10$, charge $\tilde{q}=0.1$ and
for varying values of correction parameter $\varrho$ .
At a peak value the temperature attains a maximum height and then it drops down gradually and obtain a condition of asymptotically flat by indicating the stability of BH as $r_{+}\rightarrow\infty$. It can be observed that the $T'_{H}$ decreases as we increase the correction parameter values.
The temperature of BH increases with the decreasing values of event horizon. This physical presentation reflects the stability state of BH. The maximum temperature at non-zero horizon left the BH remnant.

\textbf{Figure 3(ii)} represents the behavior of $T'_{H}$ via $r_{+}$ with fixed values of mass $m=1$, correction parameter $\varrho=0.8$, charge $\tilde{q}=0.1$, free parameters $\tilde{a}_1=1, \tilde{b}_1=0.1, \tilde{a}_2=50, \tilde{b}_2=-10$ and for varying values of $a$.
There can be seen that for different values of $a$ the corrected temperature gets a height and then it shows an asymptotically flat behavior. It is notable that when we increase the value of $a$ the corrected temperature decreases as well as for the increasing value of horizon the corrected temperature also decreases. This Hawking's phenomenon represents the BH stable condition in the domain $0 \leq r_{+} \leq 15$. From both plots, we can observe that for $T'_{H}\geq 0$, the BH gets its stable form while for $T'_{H}<0$ the BH with negative temperature always depicts its unstable form. We can also observe it graphically that the $T'_{H}$ is less than the original one. So, we can conclude the quantum corrections decelerates the increment in temperature.

\begin{center}
\includegraphics[width=7cm]{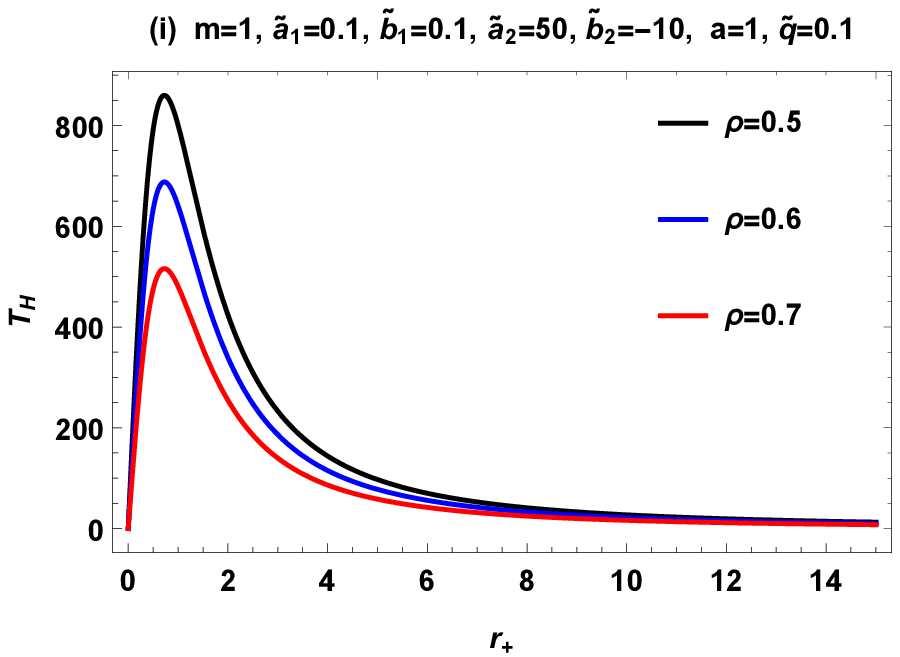}\includegraphics[width=7cm]{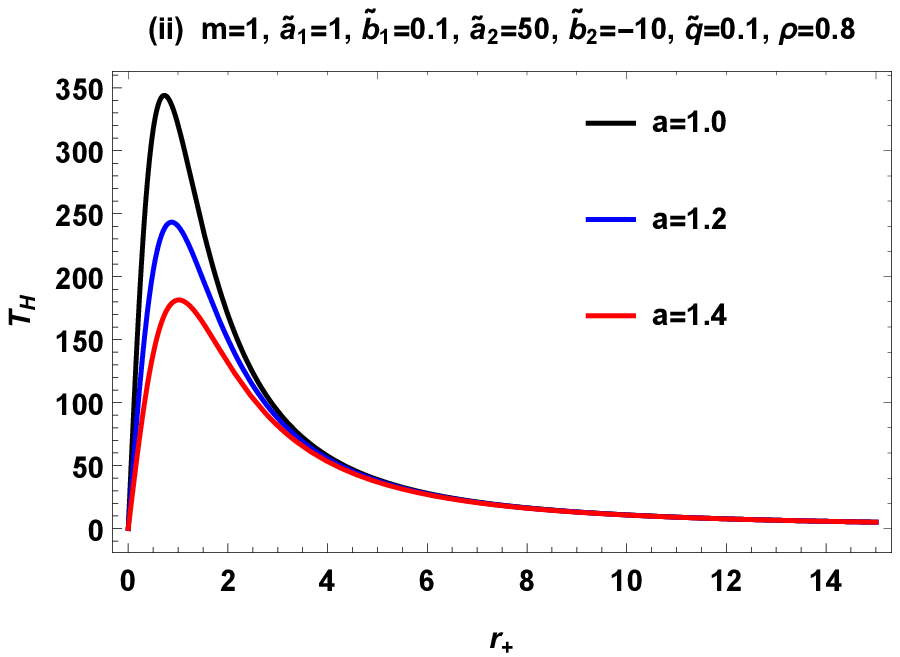}\\
{Figure 3: $T'_{H}$ versus $r_{+}$ with $\Xi=1$.}
\end{center}

\section{Summary and Discussion}
The theory of null Aether is a vector-tensor gravity theory with null vector and Aether field exist at every point of the spacetime. In this paper, we have studied a new asymptotically flat BH solution by using Newman-Janis algorithm.
To do so, firstly, we have reviewed the asymptotically flat BH solution in NAT and then by applying the Newman-Janis algorithm, we have derived a new asymptotically flat NAT BH spacetime influenced by rotation parameter.
By considering the spin parameter $(a \rightarrow 0)$ in Eq. (\ref{k1}), we get the asymptotically flat BH solution \cite{1a} in general relativity.
Furthermore, by taking into account the surface gravity $\kappa$, we have computed the temperature for NAT BH in the presence of rotation parameter. The BH temperature depends upon the charge, mass, spin and free parameters of the BH.
The NAT BH temperature in Eq. (\ref{11}) recovers the temperature of Schwarzschild BH for $\tilde{q}=0=a$ as in Eq. (\ref{m1}).
Moreover, we have comprised the graphical representation of Hawking temperature w.r.t event horizon in order to check the stability of BH.

We have studied the radiation spectrum through bosonic tunneling process of spin-$1$ particles from NAT BH involving both spin and quantum gravity parameters. Therefore, we have utilized the generalized Lagrangian equation incorporating the GUP influenced by quantum gravity. For this investigation, we have applied the Hamilton-Jacobi ansatz and WKB approximation to the generalized Lagrangian field equation for boson particles. We have obtained the bosonic corrected tunneling rate of emitted particles and their corresponding corrected temperature $T'_{H}$. It has note worthy to analyzed that, when we ignore the quantum gravity effects, i.e., ($\rho=0$), then the corrected Hawking temperature in Eq. (\ref{thh}) is reduced to the original temperature in Eq. (\ref{11}).
The corrected temperature of BH depends upon spin parameter, quantum gravity parameter and Aether field. The $T'_{H}$ reduces into Schwarzschild BH temperature when the spin parameter, quantum gravity parameter and Aether field approaches to zero. It has been analyzed that the quantum gravity decelerates the increase in $T'_{H}$ in the process of radiation. Moreover, we have analyzed the physical significance of corrected temperature to check the effects of quantum gravity and rotation parameter on $T'_{H}$ by seeing the stability of NAT BH over Aether field.
The results from the plots of Hawking temperature with respect to the horizon in the presence/absence of gravity parameter for the given BH are given as follows:

\begin{itemize}
    \item In the absence of gravity parameter the temperature shows the asymptotically flat behavior in the range of charge $0.1\leq\tilde{q}\leq 0.3$ and the $T_{H}$ decreases with the increasing $r_{+}$.  This is physical graphical presentation of $T_{H}$ w.r.t $r_{+}$ and depicts the stable condition of BH with positive temperature.
    \item The $T_H$ for varying values of rotation parameter $a$ shows an asymptotically flat behavior and after a maximum height the temperature goes on decreasing as well as for the increasing horizon. This Hawking's phenomenon depicts the BH stability in the domain $0\leq r_{+}\leq 15$.

    \item In the presence of gravity parameter $T'_{H}$ decreases with the increasing values of correction parameter as well as horizon. We have observed BH remnant at nonzero horizon with maximum temperature for different values $\rho$ in the domain $0\leq r_+\leq 15$.
    \item For different values of $a$ the corrected temperature gets a height and then it shows an asymptotically flat behavior. It is notable that the corrected temperature decreases with the increasing values of $a$ as well as for the increasing value of horizon. This Hawking's phenomenon represents the BH stable condition in the domain $0\leq r_{+}\leq 15$.
    \item From all the plots, we have observed that for $T'_{H}\geq 0$, the BH gets its stable form. We have also observed it graphically that the $T'_{H}$ is less than the original one. So, we have concluded that the quantum corrections decelerates the increment in temperature.
\end{itemize}
\section{Appendix}
After setting the all values in Eq. (\ref{xx}), we get the field equations set as
\begin{eqnarray}
&&\frac{K}{G(FK+L^2)}\Big[c_{1}(\partial_{0}Q_{0})(\partial_{1}Q_{0})+\varrho c_{1}
(\partial_{0}Q_{0})^{3}(\partial_{1}Q_{0})-c_{0}(\partial_{1}Q_{0})^{2}
-\varrho c_{0}(\partial_{1}Q_{0})^4+c_{1}eA_{0}(\partial_{1}Q_{0})\nonumber\\
&&+c_{1}\varrho eA_{0}(\partial_{0}Q_{0})^{2}(\partial_{1}Q_{0})\Big]
-\frac{L}{G(FK+L^2)}\Big[c_{3}(\partial_{1}Q_{0})^2+\varrho c_{3}(\partial_{1}
Q_{0})^4-c_{1}(\partial_{1}Q_{0})(\partial_{3}Q_{0})-\varrho c_{1}
(\partial_{1}Q_{0})(\partial_{3}Q_{0})^2\Big]\nonumber\\
&&+\frac{K}{H(FK+L^2)}\Big[c_{2}(\partial_{0}Q_{0})(\partial_{2}Q_{0})
+\varrho c_{2}(\partial_{0}Q_{0})^3(\partial_{2}Q_{0})-c_{0}(\partial_{2}
Q_{0})^2-\varrho c_{0}(\partial_{2}Q_{0})^4+c_{2}eA_{0}(\partial_{2}Q_{0})
+c_{2}eA_{0}\varrho\nonumber\\&&(\partial_{0}Q_{0})^{2}(\partial_{1}Q_{0})\Big]
+\frac{FK}{(FK+L^2)^2}\Big[c_{3}(\partial_{0}Q_{0})(\partial_{3}Q_{0})
+\varrho c_{3}(\partial_{0}Q_{0})^{3}(\partial_{3}Q_{0})-c_{0}(\partial_{3}Q_{0})^{2}-\varrho c_{0}(\partial_{3}Q_{0})^4+c_{3}eA_{0}\nonumber\\&&(\partial_{3}Q_{0})+c_{3}eA_{0}
(\partial_{0}Q_{0})^{2}(\partial_{3}Q_{0})\Big]-m^2\frac{K c_{0}-L c_{3}}{(FK+L^2)}=0,\label{jj}\\
&&\frac{-K}{G(FK+L^2)}\Big[c_{1}(\partial_{0}Q_{0})^2+\varrho c_{1}
(\partial_{0}Q_{0})^4-c_{0}(\partial_{0}Q_{0})(\partial_{1}Q_{0})
-\varrho c_{0}(\partial_{0}Q_{0})(\partial_{1}Q_{0})^{3}
+c_{1}eA_{0}(\partial_{0}Q_{0})\nonumber\\
&&+\varrho c_{1}eA_{0}(\partial_{0}Q_{0})^3\Big]+\frac{L}{G(FK+L^2)}
\Big[c_{3}(\partial_{0}Q_{0})(\partial_{1}Q_{0})+\varrho c_{3}
(\partial_{0}Q_{0})(\partial_{1}Q_{0})^3-c_{1}(\partial_{0}Q_{0})(\partial_{3}Q_{0})-\varrho c_{1}(\partial_{0}Q_{0})(\partial_{3}Q_{0})^{3}\Big]\nonumber\\
&&+\frac{1}{GH}\Big[c_{2}(\partial_{1}Q_{0})(\partial_{2}Q_{0})
+\varrho c_{2}(\partial_{1}Q_{0})(\partial_{2}Q_{0})^3-c_{1}(\partial_{2}Q_{0})^{2}-\varrho c_{1}(\partial_{2}Q_{0})^{4}\Big]+\frac{1}{G(FK+L^2)}\Big[c_{3}
(\partial_{1}Q_{0})(\partial_{3}Q_{0})+\varrho c_{3}\nonumber\\
&&(\partial_{1}Q_{0})(\partial_{3}Q_{0})^3-c_{1}(\partial_{3}Q_{0})^2-\varrho c_{1} (\partial_{3}Q_{0})^{4}\Big]+\frac{eA_{0}K}{G(FK+L^2)}\Big[c_{1}
(\partial_{0}Q_{0})+\varrho c_{1}(\partial_{0}Q_{0})^3
-c_{0}(\partial_{1}Q_{0})-\varrho c_{0}(\partial_{1}Q_{0})^3\nonumber\\
&&+eA_{0}c_{1}+\varrho c_{1}eA_{0}(\partial_{0}Q_{0})^{2})\Big]
+\frac{eA_{0}L}{G(FK+L^2)}\Big[c_{3}(\partial_{1}Q_{0})+\varrho c_{3}(\partial_{1}Q_{0})^3
-c_{1}(\partial_{3}Q_{0})-\varrho c_{1}(\partial_{1}Q_{0})^3\Big]-\frac{m^2 c_{1}}{G}=0,\\
&&\frac{K}{H(FK+L^2)}\Big[c_{2}(\partial_{0}Q_{0})^2+\varrho c_{2}
(\partial_{0}Q_{0})^{4}-c_{0}(\partial_{0}Q_{0})(\partial_{2}Q_{0})
-\varrho c_{0}(\partial_{0}Q_{0})(\partial_{2}Q_{0})^3
+c_{2}eA_{0}(\partial_{0}Q_{0})+\varrho c_{2}eA_{0}(\partial_{0}Q_{0})^{3}\Big]\nonumber\\
&&+\frac{1}{GH}\Big[c_{2}(\partial_{1}Q_{0})^2+\varrho c_{2}
(\partial_{1}Q_{0})^{4}-c_{1}(\partial_{1}Q_{0})(\partial_{2}Q_{0})
-\varrho c_{1}(\partial_{1}Q_{0})(\partial_{2}Q_{0})^3\Big]-\frac{L}{H(FK+L^2)}
\Big[c_{2}(\partial_{0}Q_{0})(\partial_{3}Q_{0})\nonumber\\
&&+\varrho c_{2}(\partial_{0}Q_{0})^{3}
(\partial_{3}Q_{0})-c_{0}(\partial_{0}Q_{0})(\partial_{3}Q_{0})
-\varrho c_{0}(\partial_{0}Q_{0})^3 (\partial_{3}Q_{0})+c_{2}eA_{0}(\partial_{3}Q_{0})
+\varrho c_{2}eA_{0}(\partial_{3}Q_{0})^{3}\Big]\nonumber\\
&&+\frac{F}{H(FK+L^2)}\Big[c_{3}(\partial_{2}Q_{0})(\partial_{3}Q_{0})+\varrho c_{3}
(\partial_{2}Q_{0})^{3}(\partial_{3}Q_{0})-c_{2}(\partial_{3}Q_{0})^2
-\varrho c_{2}(\partial_{3}Q_{0})^4\Big]-\frac{m^2 c_{2}}{H}\nonumber\\
&&+\frac{eA_{0}K}{H(FK+L^2)}\Big[c_{2}(\partial_{0}Q_{0})+\varrho c_{2}
(\partial_{0}Q_{0})^3-c_{0}(\partial_{2}Q_{0})-\varrho c_{0}
(\partial_{2}Q_{0})^3+c_{2}eA_{0}+c_{2}\varrho eA_{0}(\partial_{0}Q_{0})^2\Big]=0,\\
&&\frac{FK-F^2}{(FK+L^2)^2}\Big[c_{3}(\partial_{0}Q_{0})^2+\varrho c_{3}
(\partial_{0}Q_{0})^4-c_{0}(\partial_{0}Q_{0})(\partial_{3}
Q_{0})-\varrho c_{0}(\partial_{0}Q_{0})(\partial_{3}Q_{0})^{3}
+{eA_{0}c_3}(\partial_{0}Q_{0})\nonumber\\
&&+\varrho c_{3}eA_{0}(\partial_{0}Q_{0})^{3}\Big]
-\frac{K}{H(FK+L^2)}\Big[c_{3}(\partial_{1}Q_{0})^2+\varrho c_{3}
(\partial_{1}Q_{0})^{4}-c_{1}(\partial_{1}Q_{0})(\partial_{3}Q_{0})
-\varrho c_{1}(\partial_{1}Q_{0})(\partial_{3}Q_{0})^3\Big]\nonumber\\
&&-\frac{L}{H(FK+L^2)}\Big[c_{2}(\partial_{0}Q_{0})(\partial_{2}Q_{0})+\varrho c_{2}
(\partial_{0}Q_{0})^3(\partial_{2}Q_{0})-c_{0}(\partial_{2}Q_{0})^{2}
+\varrho c_{0}(\partial_{2}Q_{0})^4+{eA_{0}c_2}(\partial_{2}Q_{0})
+\varrho c_{2}eA_{0}\nonumber\\
&&(\partial_{0}Q_{0})^{2}(\partial_{2}Q_{0})\Big]-\frac{eA_{0}F}{H(FK+L^2)}
\Big[c_{3}(\partial_{2}Q_{0})^2+\varrho c_{3}(\partial_{2}
Q_{0})^4-c_{2}(\partial_{2}Q_{0})(\partial_{3}Q_{0})
-\varrho c_{2}(\partial_{0}Q_{0})(\partial_{3}Q_{0})^{3}\Big]\nonumber\\
&&+\frac{eA_{0}FD-F^2}{(FK+L^2)^2}\Big[c_{3}(\partial_{0}Q_{0})+\varrho c_{3}(\partial_{0}Q_{0})^3
-c_{0}(\partial_{3}Q_{0})-\varrho c_{0}(\partial_{3}Q_{0})^3+c_{3}eA_{0}
+\varrho eA_{0}(\partial_{0}Q_{0})^2\Big]\nonumber\\&&-\frac{m^2Lc_{0}-Fc_{3}}{(FK+L^2)}=0.\label{uu}
\end{eqnarray}

\end{document}